\documentclass[10pt,letterpaper]{article}
\usepackage{opex3}
\usepackage[T1]{fontenc}
\usepackage{amsmath}
\usepackage{amsfonts}
\usepackage{amssymb}
\usepackage{graphicx}
\usepackage{color}
\usepackage{float}
\DeclareTextSymbol{\degre}{T1}{6}
\usepackage{epstopdf}
\DeclareGraphicsExtensions{.eps}

\begin{document}

\title{High-performance and power-efficient 2$\times$2 optical switch on Silicon-on-Insulator}

%\title{High contrast and power-efficient thermally-controlled 2 $\times$ 2 optical switch on Silicon-on-Insulator}
\author{Zheng Han$^{1,+}$, Gr\'egory Moille$^{2}$, Xavier Checoury$^1$, J\'{e}r\^{o}me Bourderionnet$^{2}$, Philippe Boucaud$^1$, Alfredo De Rossi$^{2}$ and Sylvain Combri\'e$^{2,*}$}

\address{$^1$Institut d'\'Electronique Fondamentale, Univ. Paris-Sud CNRS UMR 8622, 91405 Orsay, France\\
	$^2$Thales Research and Technology, 1 Av. Augustin Fresnel, 91767 Palaiseau, France}

\email{$^+$zheng.han@ief.u-psud.fr}
\email{$^*$sylvain.combrie@thalesgroup.com} %% email address is required

\date{\today}%

% % % % Abstract % % % %
\begin{abstract} % % Max 100 mots % %
A compact  ($15 \mu m \times 15 \mu m$) and highly-optimized $2\times2$ optical switch is demonstrated on a CMOS-compatible photonic crystal technology. On-chip insertion loss are below 1 dB, static and dynamic contrast are 40 dB and $>$20 dB respectively. Owing to efficient thermo-optic design, the power consumption is below 3 mW while the switching time is 1 $\mu$s.
\end{abstract}

% % % % % % % % % % % %
%\maketitle
%\bibliographystyle{plain}
%\bibliography{DCbiblio}

% % % % Intro % % % %
\section{Introduction}
\indent Optical switches play a crucial role in integrated photonic circuits, particularly for optical interconnects\cite{Farrington2013,Porter2013}, high-performance computers\cite{Dangel2015}, and sensors\cite{Fegadolli2015}. In modern data center, they add flexibility through the dynamic reconfiguration of the optical networks. More specifically, optical switches are used to route data, to control the power level or the time delay.
The potential for scalability of the switching fabric \cite{Nikolova2015} depend on the intrinsic performances of the core building block, the $2\times2$ switch, e.g. the including crosstalk, the insertion loses, the  power consumption, the speed and the footprint.\\
A variety of design concepts and technologies have been introduced for this device, \cite{Subbaraman2015} including optical MEMS\cite{han2015}, thermo-optic or carrier injection in directional coupler, Mach-Zenhder\cite{VanCampenhout2009,Dong2010} or micro-cavities\cite{Fegadolli2012,Wang2008}. The crucial point is to achieve both power-efficiency, dynamic contrast and compactness, still without introducing too much complexity in the fabrication process.
In this respect, the enhanced light-matter interaction in Photonic Crystals (PhC) has been exploited to reduce footprint and power consumption. Demonstrated CMOS-based PhC devices entail detectors\cite{haret2010,hayakawa2013}, modulators\cite{nguyen2012}, delay-lines\cite{ishikura2012}, correlators\cite{ishikura2011} and switches, either configured as a directional couplers \cite{zablocki2010} or a Mach-Zehnder interferometers \cite{vlasov2005,gu2007}, or  even exploiting the peculiar properties of PhCs, namely a tuneable transmission spectral gap\cite{zhao2013}. Tuning is either based on free carriers (by injection\cite{zablocki2010} or depletion) or exploits the thermo-optic effect.\\

\indent In this paper, we report on the design, the fabrication and the operation of a $2\times2$ optical switch based on a Photonic Crystal (PhC) directional coupler. The device is controlled through the thermo-optic effect, which is preferred here owing to the simplicity of the implementation, the large thermo-optic coefficient of silicon\cite{Cocorullo1999} and because it avoids free carrier absorption. The small size of the device implies a small enough thermal capacitance to allow a time response in the $\mu s$ scale.
%\\
%
%
% % % % Description de l'échantillon % % % %
\section{Device fabrication}
\begin{figure}[H]
	\centering
	\includegraphics[width=1\linewidth]{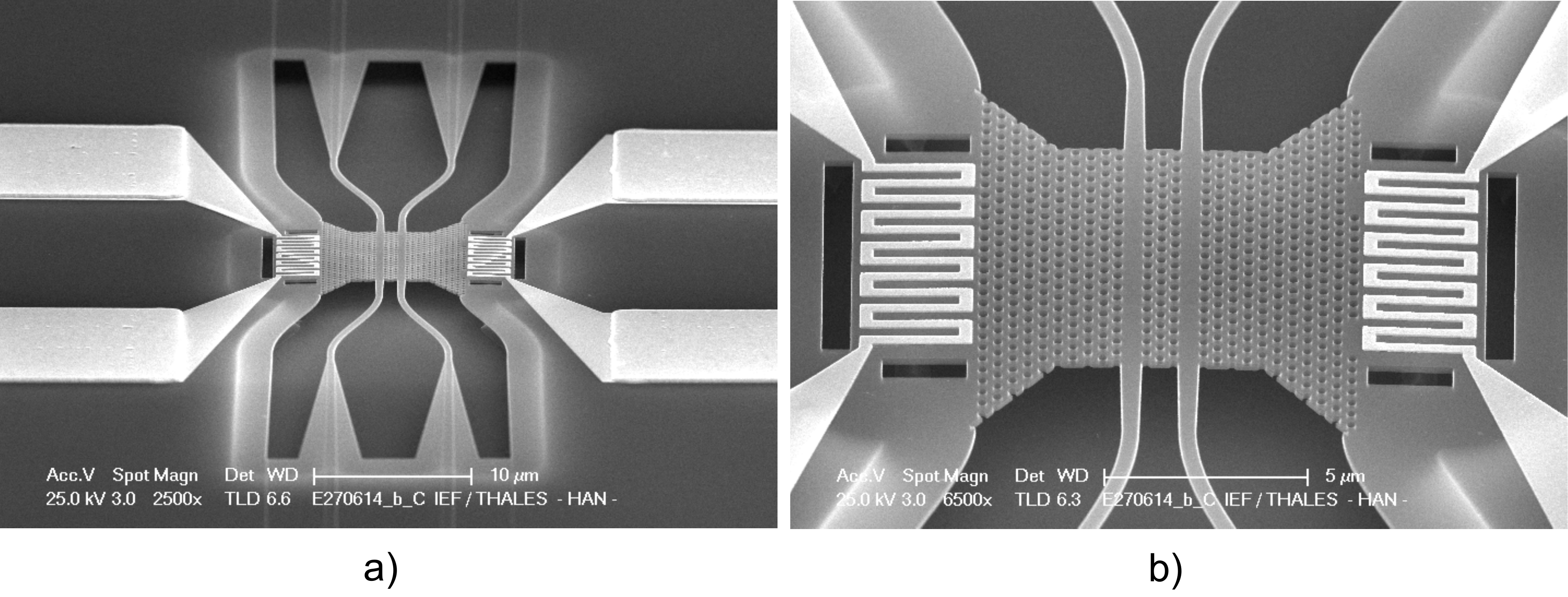}
	\caption{SEM picture of the PhC switch. a): overview of the device with the access waveguides and the electrical paths for the control. b): closer view of the PhC directional coupler.}
	\label{fig1}
\end{figure}

%\textbf{\underline{Description de l'\'{e}chantillon}}\\
%\newline
\indent The sample is fabricated on a 220nm thick Silicon-On-Insulator wafer (SOI) from SOITEC. A shallow (h=70nm) ICP-plasma etch step is first performed to form the access waveguides. Another plasma etch step down to the insulator layer forms the rib waveguides and the photonic crystal devices (Fig.\ref{fig1}a). Then, serpentine-shaped thermal heaters made of a 50nm-thick layer of platinum are deposited and patterned nearby the photonic crystal by e-beam evaporation. The electrical connections and the contact pads (not shown on the SEM picture) are then formed by deposition of a thicker layer of metal (50nm of Pt followed by 200nm of gold). After dicing with a diamond saw,  the Photonic Crystal silicon layer is transformed into an air-suspended membrane (Fig.\ref{fig1}b) by selective chemical etching of the underlying insulator by Hydrofluoric acid. The definition of the patterns has been performed by e-beam lithography (Nanobeam NB4). %The completed device is shown in figure~\ref{fig1}.
% % % % Figure 1 % % % %
%

% % % % Modélisation optique % % % %
\section{Device design}
\begin{figure} %[H]
	\centering
	\includegraphics[width=1.0\linewidth]{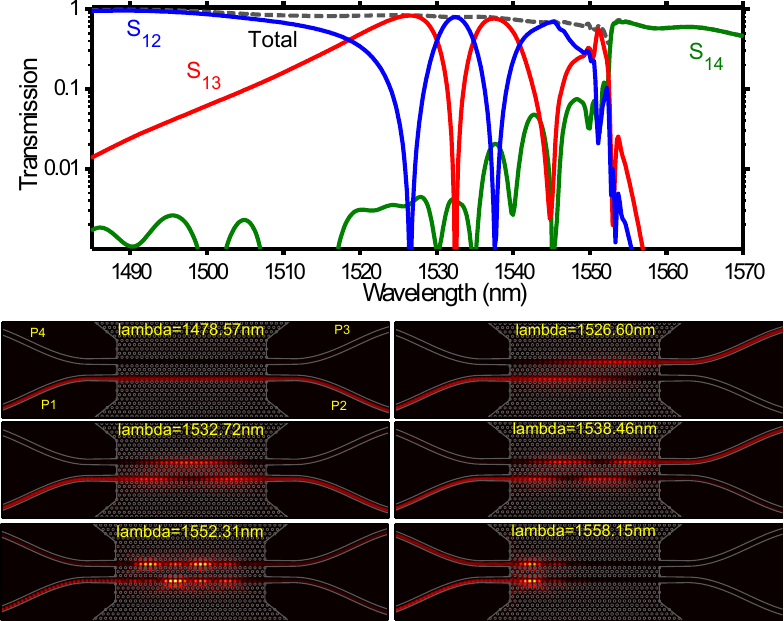}
	\caption{Top: calculated transfer function and total (FDTD). Bottom: magnetic field distribution (H-field) at some representative wavelengths. }
	\label{fig2}
\end{figure}

%\textbf{\underline{Mod\'{e}lisation optique}}\\
\indent  The design of the $2\times2$ switch is based on the well established concept of Directional Coupler, where light is periodically transferred from one waveguide to another owing to the interference of two supermodes. The spatial period is know as \textit{beat length} and it is related to the wavevectors of the even and odd supermodes: $L_b^{-1}=|\beta_e-\beta_o|$. Short devices require a large difference in the wavevector which ultimately translates into a large index contrast. This motivates the interest in high index contrast structures such as Photonic Crystals, which have been indeed used to demonstrate an optical modulator based on a Mach-Zenhder design \cite{vlasov2005,zhao2013}. Yamamoto and coworkers have pointed out the potential of PhC dispersion engineering for a compact directional coupler\cite{yamamoto_photonic_2006}, which was then implemented as an all-optical switch \cite{ofaolain_compact_2010} or a $2\times2$ electrically controlled switch \cite{zablocki2010}.\\
Our design is based on two identical waveguides created by removing one row of holes in a hexagonal lattice of holes with nominal radius equal to 0.23{\itshape a} $(${\itshape a} is the period of the PhC equal to 414nm$)$. The width of the waveguides is reduced to {\itshape a}$0.9\sqrt{3}$ bringing closer the blocs of Photonic Crystal. Then the two first rows of holes are shifted outwards by 0.1{\itshape a}. The two waveguides are separated by three rows of holes. This design allows a spectral window (Fig.\ref{fig2}), located approximately between 1525 and 1550 nm (i.e. 25 nm) where the wavevector difference $\beta_e-\beta_o$  is large and almost constant such that the corresponding beat length varies from about 40 periods to about 10, e.g 8 $\mu m$ (Fig.\ref{fig2}). Consequently, the length of the directional coupler is set to 20 periods in order to allow for a complete switching cycle as the wavelength is tuned by about 10 nm. 5 periods long sections of tapered PhC waveguides are added on both sides to ensure mode matching to the suspended photonic wires and thereby minimizing the insertion losses, particularly in the slow light regime of the PhC. The \textit{S} parameters are calculated by in-house 3D FDTD ({\itshape Finite Difference in Time Domain}) code and are shown in figure~\ref{fig2}.\\
The transfer functions reveal the dispersive nature of the beating length of the directional coupler enabling the power to go alternatively from one arm to the other one. Below 1510nm the light propagates straight from port P1 to P2, as shown on the calculated field distribution. The first maximum of the cross state ($S_{13}$) is at 1526.6 nm. At 1532.7 nm a round-trip is done. At 1552.31 nm, slow-light modes of the PhC waveguides are exited leading to input reflection and quasi-equal-power distribution in the other arms. From this, the cut-off of the PhC waveguides is reached and at 1558nm, only the slow modes of the access waveguides are excited leading to efficient power transfer form port P1 to port P4. The calculated insertion losses are due to the residual mode mismatch and are less than 1 dB around 1530 nm and they increase to a maximum of 2 dB close to the transmission edge at 1550nm (Fig.~\ref{fig2}), the propagation losses are totally negligible due to their short length. The calculated extinction is larger than 30 dB.\\
We point out that the choice of a suspended membrane design, instead of a solid substrate, is motivated by a much less constrained design and by the much better thermal insulation, which considerably reduces the power budget for tuning. In this respect, trenches (Fig. \ref{fig1}) have been added to further improve the thermal resistance. The $2\times2$ switch is meant to be connected to the other devices or switches through low-loss shallow waveguides; their mode is adiabatically converted \cite{Dong2010} into a photonic wire mode under which the silica substrate is removed before it connects to the photonic crystal (Fig. \ref{fig1}). The device layout ensures a symmetric path for the 4 ports.\\
%
% % % % Mesures % % % %
\section{Optical Characterization}
\begin{figure}[H]
	\centering
	\includegraphics[width=0.6\linewidth]{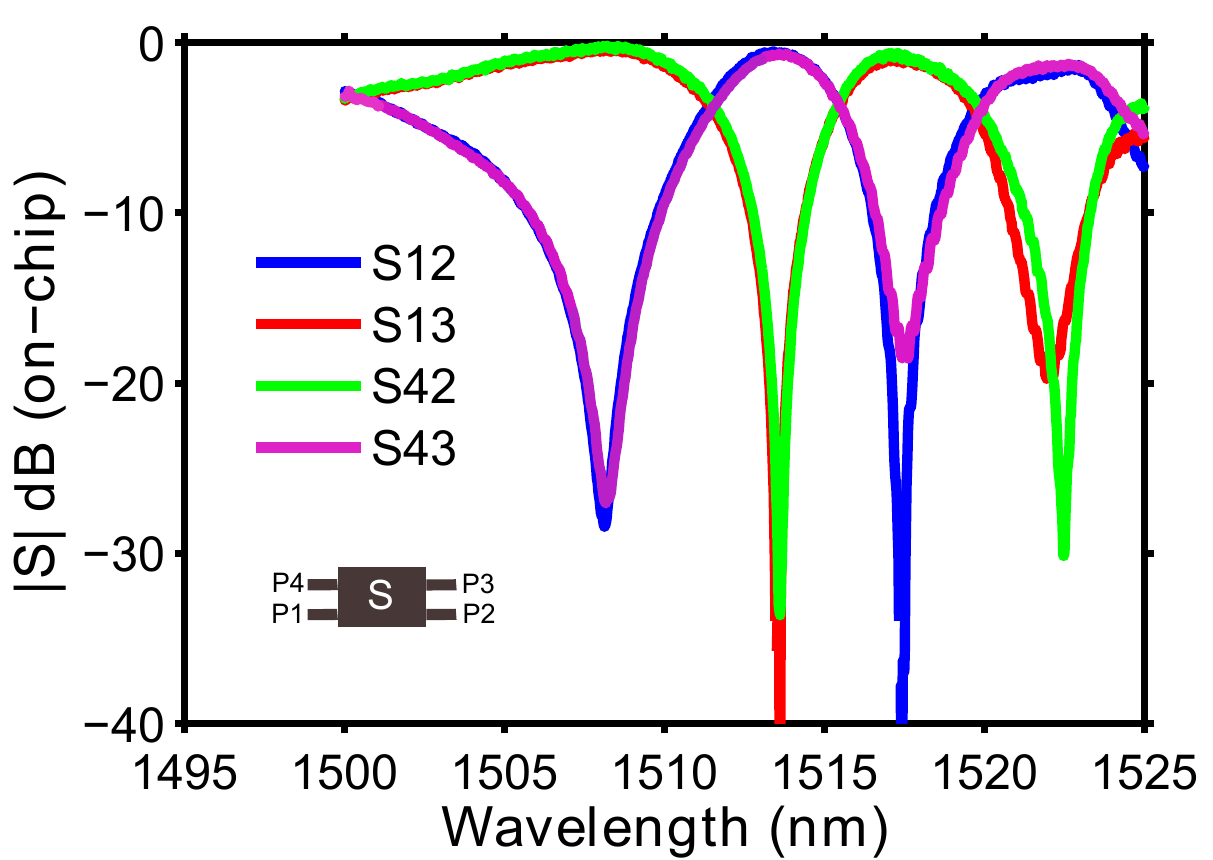}
	\caption{On-chip transmission spectra ($S$ parameters) through the four ports. Estimated insertion losses (transmission maximum) is at $-0.5$ dB.}
	\label{fig3}
\end{figure}

%\textbf{\underline{Mesures statiques}}\\
\indent The scattering parameters \textit{S} have been measured using a tunable laser, coupled to the TE mode of the shallow-etched access waveguide using a lensed fiber. Insertion losses are reduced to about 3 dB owing to an inverse taper mode adapter. The experimental spectra characterizing the $2\times2$ device are shown in Fig.~\ref{fig3}, namely $S_{12}$ and $S_{43}$ ( bar mode) and $S_{13}$ and $S_{42}$ (cross mode). The S parameters are evaluated by subtracting the off-chip losses (due to lensed fibers, access waveguides, etc.) which amount to about 6 dB, as  estimated from the measurement of a reference shallow waveguide, and which are still dominated by fiber to waveguide coupling.
The experimental $S$ parameters are in striking close correspondence to calculations, except for a 20 nm spectral offset related to the fabrication process which can be easily corrected by calibration. The degree of symmetry achieved in the fabricated device is particularly good, when comparing $S_{12}$ to $S_{43}$ and $S_{13}$ to $S_{42}$, which nearly superimpose. The extinction between minima and maxima of the spectra is way larger than 20 dB and reaches 45dB at 1514nm. This is very close to the record value achieved with a cascaded Mach-Zehnder architecture\cite{Suzuki2015}, which is more complex.
%
% % % % Mesures paramètres S % % % %
%
%
% % % % Modélisation thermique % % % %
\section{Thermal dynamics}
%\textbf{\underline{Mod\'{e}lisation thermique}}\\
\indent The switching dynamics of the device is understood by modeling the thermal dynamics\cite{tinker2005,gu2007}. This was performed using the Finite Element Method (FEM) implementation of the Fourier equation, available with the COMSOL software platform. This enables the calculation of the dynamical evolution of the temperature distribution, and therefore, to estimate the power dissipation and transient response. The thermal conductivity of air was neglected, because orders of magnitude smaller than the thermal conductivity of Silicon, while Dirichlet boundary conditions (perfect thermal conductor) have been implemented to terminate the SOI layer with a thermal sink.

\begin{figure}[H]
	\centering
	\includegraphics[width=0.7\linewidth]{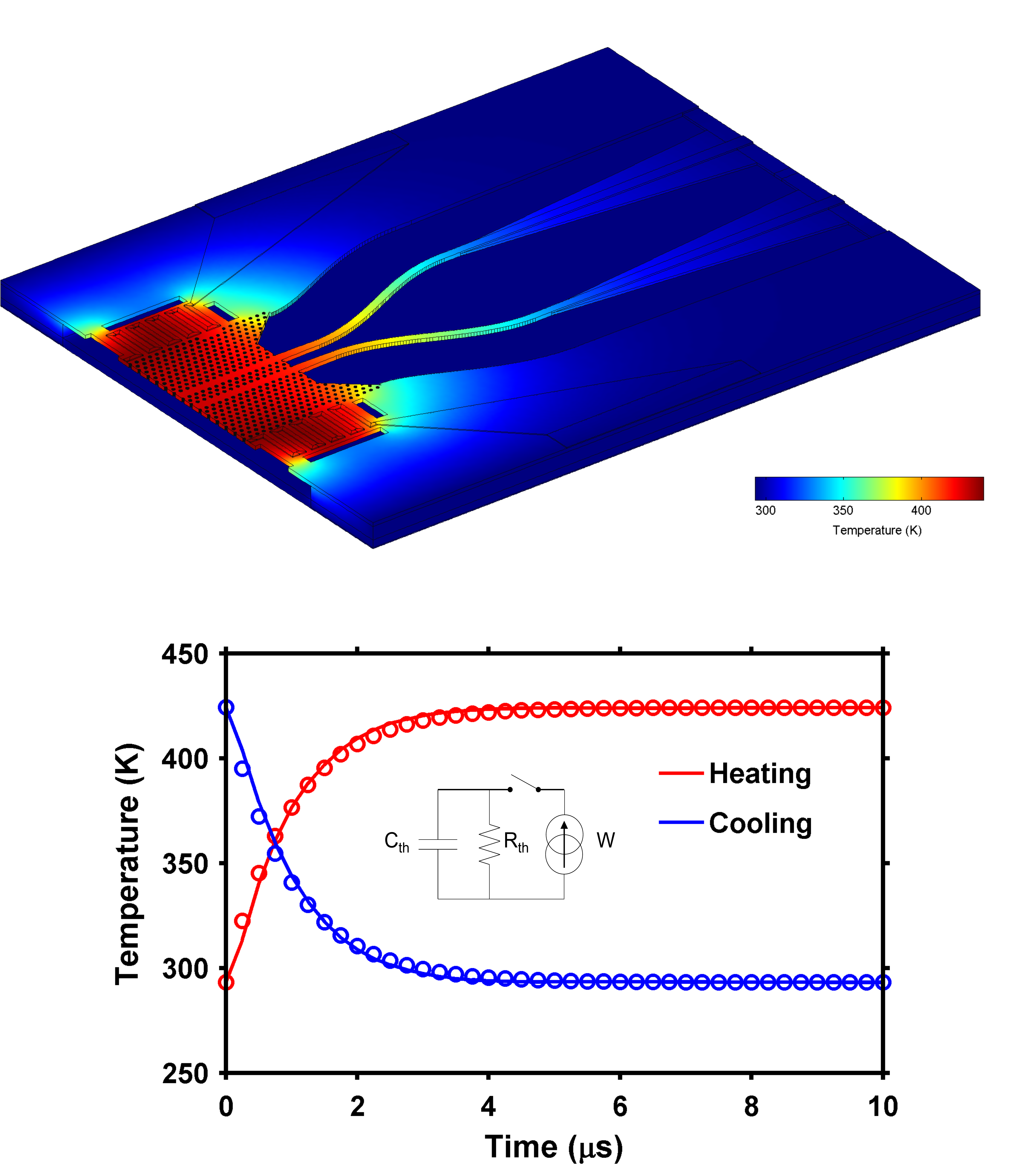}
	\caption{Top: Calculated temperature distribution at steady state. Bottom: time evolution of the temperature averaged over the PhC waveguide.}
	\label{fig4}
\end{figure}

Heat is injected by simulating the Joule dissipation of an electric current of 4 mA through the metal serpentine with 50nm x 260nm cross section (overall resistance is $225\Omega$).\\
The temperature distribution at steady state is shown in figure~\ref{fig4}. As expected, the heat is confined in the PhC area with a rather uniform distribution owing to the air-trenches. We point out that the metal serpentine itself does not contribute itself much to the heat transport, thereby reducing leaks. This is because the metal layer is very thin.\\
\indent The transient evolution of the temperature averaged along the waveguide is shown in figure~\ref{fig4}. The warming up is simulated with a current step applied to the serpentine.  The thermal relaxation is modeled taking the temperature distribution achieved at the steady state as initial condition. The corresponding two time constants extracted, $\tau_{heat}=0.99\mu s$ and $\tau_{cool}=1.01\mu s$, are very close, as expected from the idealized $RC$ model. The frequency cut-off of the switch is $f_{c}=\frac{1}{2\pi\tau}\simeq160kHz$. The device can be described by an equivalent $RC$ circuit based on the equation: $C_{th}\left(dT/dt\right)=W-T/R_{th}$, with $C_{th}$ the thermal capacitance, $W$ the power of the heating source and $R_{th}$ the thermal resistivity. Thus, the initial derivative of the temperature gives a thermal capacitance of $C_{th}=2.84\times10^{-11}J K^{-1} $ and from the steady state we can estimate the thermal resistance to be $R_{th}=3.47\times10^{4}K s J^{-1}$.
%
%
% % % % Accordabilité thermique % % % %
\section{Operation of the  $2\times2$ switch}
\begin{figure}[H]
	\centering
	\includegraphics[width=0.7\linewidth]{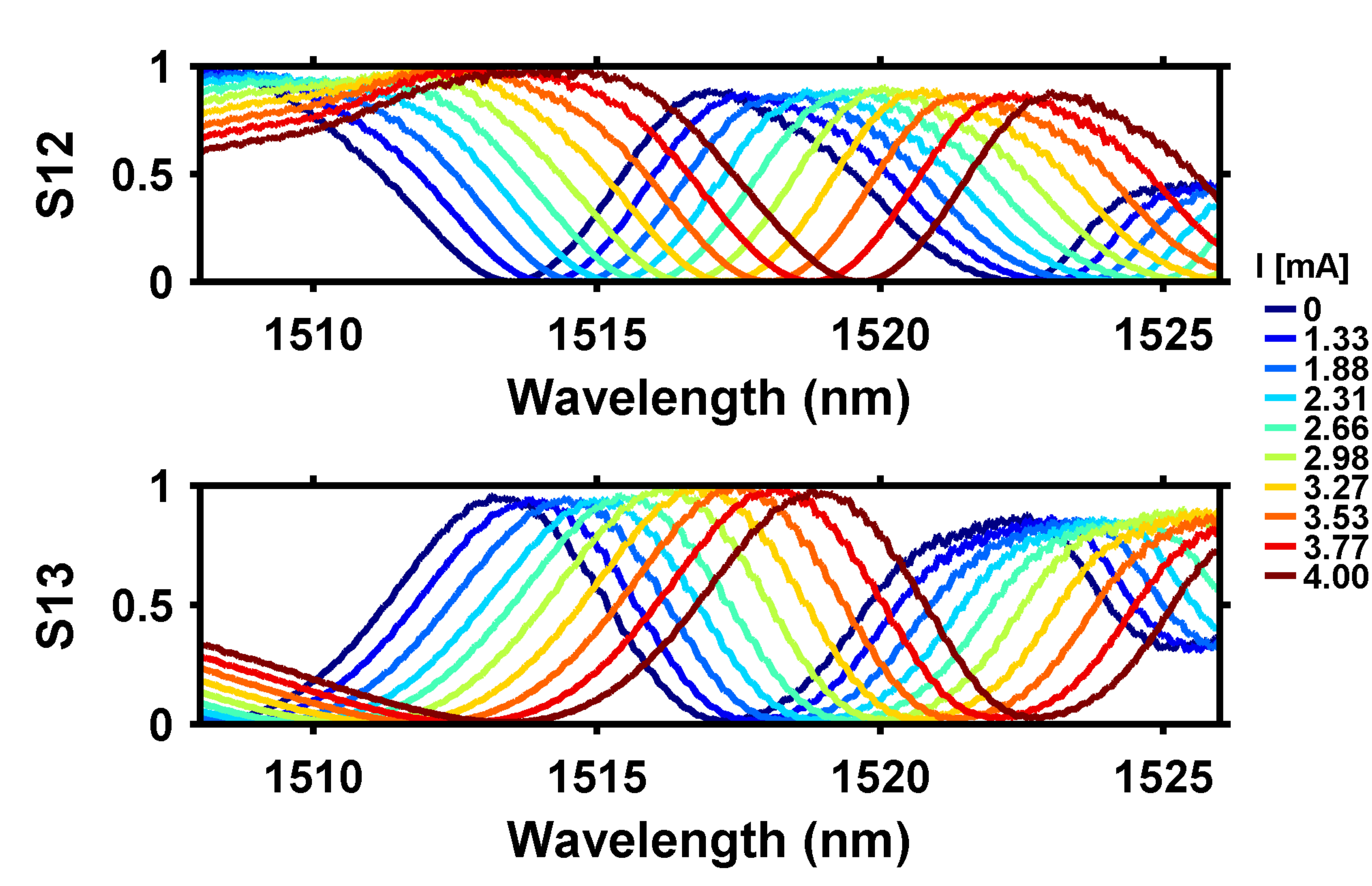}
	\caption{Experimental transmission spectra (bar $(S_{12})$ and cross $(S_{13})$) as a function of the electric control. The current is varied with a square root law to produce constant step in dissipated heat.}
	\label{fig5}
\end{figure}

%\textbf{\underline{Accordabilit\'{e} thermique}}\\
\indent
The \textit{static} characterisation of the switch is performed by measuring representative spectral transmission (bar $(S_{12})$ and cross $(S_{13})$)  at different level of the injected current. This is shown in Fig.~\ref{fig5}. The increase of the current from 0 to 4 mA over $310\Omega$ (the dissipated power raises up to 5 mW) produces an uniform displacement of the whole transmission spectra by about 7 nm. Operating around $\lambda=1517nm$, a complete transition from bar to cross is achieved by a spectral shift of $3.5$ nm, reached at a current level of 3.2 mA (hence about 3 mW of dissipated power). The corresponding temperature rise is estimated to about 42\degre C based on the relationship $\frac{d\lambda}{dT}=\frac{\partial\lambda}{\partial n}\frac{\partial n}{\partial T}$, with the silicon temperature coefficient (\cite{li1980,Cocorullo1999}) $\frac{\partial n}{\partial T}=2\times10^{-4}K^{-1}$ and the wavelength sensitivity $\frac{\partial\lambda}{\partial n}=4.2\times10^{-7}m$, as calculated by FDTD.\\

\begin{figure}[H]
	\centering
	\includegraphics[width=0.8\linewidth]{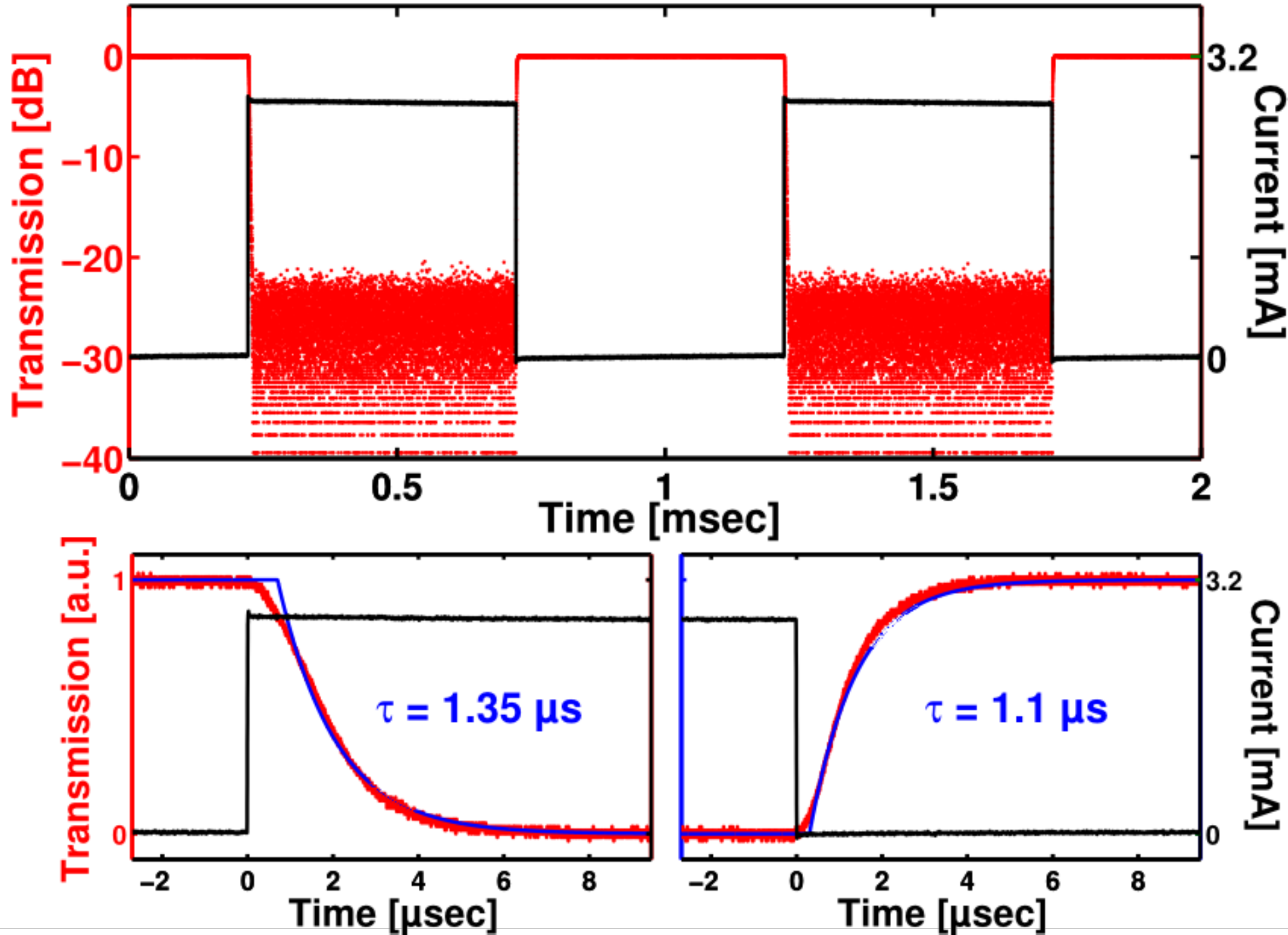}
	\caption{Transient analysis, oscilloscope traces. Top: Square-wave electric control and transmission ($S_{12}$) in logarithmic scale. Bottom: enlarged view of the fall (left) and rise (right) edge with the exponential fit.}
	\label{fig6}
\end{figure}

At last, we measure the dynamic response of the switch. The wavelength is set to a maxima of $S12$ at zero current. A square wave electric signal (0.8V and 1kHz repetition frequency), is applied to the thermal heaters. The oscilloscope trace of the detected transmission ($S12$)  is shown in figure~\ref{fig6}. It is apparent that the switch operates between two perfectly stable states, with no noticeable drift in transmission over times which are 3 orders of magnitude longer than the characteristic switching time. We estimate the dynamic contrast to exceed 20dB (here limited by the amplitude resolution of the oscilloscope).
 The switching bandwidth of the device is inferred from the falling and rising edges through an exponential fit (Fig. \ref{fig6}). The rise time, $\tau=1.1\mu s$, is very close to calculations, while the fall time, $\tau=1.35\mu s$, is slightly longer.
Thus,  the frequency cut-off of the device is about 130kHz, which is remarkably fast when considering the thermal origin of the switching mechanism. This is directly related to the small size of the device. While not as fast as carrier injection, thermal effect is however fast enough for numerous applications in particular for data centers interconnects\cite{Farrington2013} and has the advantage of being much easier to deploy and being exempt from free carriers induced absorption and dispersion.

%
% % % % Conclusion % % % %
\section{Conclusion}
%\textbf{\underline{Conclusion}}\\
\indent In conclusion we have demonstrated a high-performance and compact SOI 2 x 2 optical switch with a footprint of about $15\times15$ microns (including rib-access waveguides and integrated heaters), less than 1 dB on chip losses, while total insertion losses (fiber-to-fiber total are also competitive (6dB). The very good symmetry of the response over the 4 ports reveals an excellent mastery of the fabrication and the large  dynamic contrast (>20dB) is achieved with a maximum large-signal modulation bandwidth of 130kHz. The power budget is extremely competitive, as a complete transition from cross to bar is achieved with about 3 mW  of electric power. A quick extrapolation of these figures to a large switching matrix promises for very competitive devices.
\section*{Acknowledgments}
\indent This work was supported by the French National Research Agency under the ``Symphonie'' project. The authors acknowledge the support from the French "RENATECH" network.\\
\bibliographystyle{ieeetr}
\bibliography{DCbiblio}

\end{document}